\documentclass[twoside]{article}
\usepackage[a4paper]{geometry}
\usepackage[latin1]{inputenc} 
\usepackage[T1]{fontenc} 
\usepackage{RR}
\usepackage{hyperref}
\usepackage{amssymb,amsmath} 
\usepackage{appendix}
\usepackage{subfigure}
\usepackage{tikz}
\RRNo{8650}
\RRdate{December 2014}
\RRauthor{
  Pierre Brunisholz\thanks{Grenoble INP, LIG, CNRS UMR 5217, Grenoble, France}
  \and
Marine Minier\thanks{Université de Lyon, INRIA Privatics,  
  INSA-Lyon, CITI-INRIA, F-69621, Villeurbanne, France}%
\and Fabrice Valois\thanks{Université de Lyon, INRIA Urbanet,  
  INSA-Lyon, CITI-INRIA, F-69621, Villeurbanne, France}}
\authorhead{Brunisholz \& Minier \& Valois}

\RRtitle{Apport du codage réseau pour le routage dans les réseaux de capteurs sans fil
}
\RRetitle{The Gain of Network Coding in Wireless Sensor Networking}
\titlehead{The Gain of Network Coding in Wireless Sensor Networking}
\RRresume{Les Réseaux de Capteurs Sans Fils ont des caractéristiques bien connues telles que la faible consommation énergétique, la prise en compte des changements de topologie, le déploiement en milieu ouvert, les liens radio peu fiable, etc.
	Dans ce rapport, nous proposons de montrer les intérêts liés à l'introduction de Codage Réseau au sein des Réseaux de Capteurs Sans Fils, notamment la résilience.

L'une de nos préoccupations concerne la résilience des Réseaux de Capteurs Sans Fils.
	Il a été montré que la résilience pouvait être représentée comme une métrique à plusieurs dimensions~\cite{5478822,erdene2011enhancing,6423640} ayant pour paramètres le Taux de Livraison Moyen, le Délai Moyen, l'Efficacité \'Energétique, le Débit Moyen et la Distribution des Livraisons.
	La résilience peut être représentée sous forme d'un diagramme de Kiviat ayant pour branches les précédents paramètres.
	Afin de présenter la résilience, des travaux antérieurs ont été effectués sur le protocole Random Gradient Based Routing, qui présente une bonne résilience en environnement malveillant.
	Nous voulons observer les améliorations en terme de résilience que l'introduction de Codage Réseau sur le précédent protocole apporte en fonction du nombre de n\oe{}uds malveillants. 
}
\RRabstract{	Wireless Sensor Networks have some well known features such as low battery consumption, changing topology awareness, open environment, non reliable radio links, etc.
	In this paper, we investigate the benefits of Network Coding Wireless Sensor networking, especially resiliency.

One of our main concern is the resiliency in Wireless Sensor Networks.
	We have seen that resiliency could be described as a multi dimensional metric~\cite{5478822,erdene2011enhancing,6423640} taking parameters such as Average Delivery Ratio, Delay Efficiency, Energy Efficiency, Average Throughput and Delivery Fairness into account.
	Resiliency can then be graphically represented as a kiviat diagram created by the previous weighted parameters.
	In order to introduce these metrics, previous works have been leaded on the Random Gradient Based Routing, which proved good resiliency in malicious environment.
	We look for seeing the improvements in term of resiliency, when adding network coding in the Random Gradient Based Routing with malicious nodes.
	}
\RRmotcle{Réseau de capteurs sans fil, routage, codage réseau, résilience}
\RRkeyword{Wireless Sensor Netowrks, routing protocols, network coding, resiliency}
\RRprojets{Urbanet and Privatics}
 \URRhoneAlpes 

\begin{document}
\makeRR   

\section{Introduction}
Wireless Sensor Networks have two critical issues: energy consumption and delivery efficiency.

Because sensors are designed to be used in noisy environments for a reasonably long time, and because they are small sized, which mean they cannot embed a large capacity battery when they are not just harvesting energy, increasing lifetime is still a challenge.
In wireless networking, the energy dissipation comes from transmission.
Illustrative, in terms of energy, it is often said that sending one bit over the radio channel is equal to one thousand processor cycles~\cite{Karl:2005:PAW:1076303}.
That is why it looks important to optimize the messages sensors have to send each other.
Most of the traffic is generated by packets used to ensure the proper functioning of the network: beacons and control packets.
While it has been shown that beacons could be avoided by trading them with the node's neighbors position knowledge, there are a lot of control packets like \textit{ACK, Interest Messages, Route Request, etc.} in use.

On the other hand, due to the nature of wireless communications, delivery efficiency is a real challenge.
In practice, the radio link is not reliable~\cite{Karl:2005:PAW:1076303}, it is asymmetric, and because sensors are deployed in open environments, there is a lot of collisions, interferences, and possible malicious entities.
That is why there is a need in resiliency improvement.

In order to solve these two majors challenges, we think about using Network Coding, and study the impact on them.
Network Coding consists in creating packets containing linear dependencies with other packets.
In other words, it creates message containing a little information from different messages.

This leads us to study the impact of network coding in resiliency, in malicious environment.

This paper is organized as follows.
Section~\ref{ScientificBackground} gives a general introduction to Network Coding, introduces some basic routing protocols and how it  is possible to combine them with Network Coding. In Section~\ref{Resiliency} we will study the impact of network coding on resiliency in malicious environment, with a Random Gradient Based Routing protocol.
Finally we will summarize the results and present further works in Section~\ref{Conclusion}.


\section{Related Work}\label{ScientificBackground}
\subsection{Network Coding}
Tracey Ho and Desmond Lun gave several definitions of Network Coding~\cite{Ho:2008:NCI:1817176}.
Basically, we can say that \textit{Coding at a node in a network is Network Coding}, where coding means a causal mapping from inputs to outputs.
This definition has the inconvenient of not distinguishing the network coding we are going to speak about, from the channel coding used in noisy networks.

We will then define the Network Coding as \textit{coding at a node in a network with error-free links}.
Moreover, this definition helps us to make a difference between Network Coding and source coding.

But this definition can be more specific, and if we are considering that we are in packets networks, we can define Network Coding as \textit{coding content of packets inside a node}.
If we had a little generalisation by saying that we apply the coding above the physical layer, we can distinguish the Network Coding function from the information theory.
Then we base our work on the previous definition of Network Coding.
\begin{figure}
	\centering
	\subfigure[General Approach]{
		\begin{tikzpicture}
			\begin{scope}
				\node[draw,circle,thick,minimum size=5mm] (S) at (0,0) {};
				\draw[->,>=latex] (-1,1) to node [auto] {$p_1$} (S);
				\draw[->,>=latex] (1,1) to node [auto] {$p_2$} (S);
				\draw[->,>=latex] (S) to node [auto] {$f(p_1,p_2)$} (0,-1);
			\end{scope}
			\begin{scope}[xshift=2cm]
				\node[draw,circle,thick,minimum size=5mm] (S) [label={$[P=p_1,p_2]$}] at (0,0) {};
				\draw[->,>=latex] (S) to node [auto] {$f(p_1,p_2)$} (0,-1);
			\end{scope}
		\end{tikzpicture}
	}
	\hspace{15pt}
	\subfigure[Inter Session Network Coding]{
		\begin{tikzpicture}
			\begin{scope}
				\node[draw,circle,thick,minimum size=5mm] (S) at (0,0) {};
				\draw[->,>=latex] (-1,1) to node [auto] {$p_1$} (S);
				\draw[->,>=latex] (1,1) to node [auto] {$p_2$} (S);
				\draw[->,>=latex] (S) to node [auto] {$p_3=p_1 \oplus p_2$} (0,-1);
			\end{scope}
		\end{tikzpicture}
		\label{InterFlowIllustr}
	}
	\hspace{15pt}
	\subfigure[The Butterfly Scheme]{
		\scalebox{0.5}{
			\begin{tikzpicture}
				\begin{scope}
					\node[draw,circle,thick,minimum size=8mm] (S)	at (0,0)	{s};
					\node[draw,circle,thick,minimum size=8mm] (D)	at (2,-1)	{B};
					\node[draw,circle,thick,minimum size=8mm] (G)	at (-2,-1)	{A};
					\node[draw,circle,thick,minimum size=8mm] (C1) at (0,-2)	{C};
					\node[draw,circle,thick,minimum size=8mm] (C2) at (0,-4)	{D};
					\node[draw,circle,thick,minimum size=8mm] (T1) at (-2,-5)	{$t_1$};
					\node[draw,circle,thick,minimum size=8mm] (T2) at (2,-5)	{$t_2$};

					\draw[->,>=latex]	(S) to node [auto] {$p_1$}	(G);
					\draw[->,>=latex]	(S) to node [auto] {$p_2$}	(D);
	
					\draw[->,>=latex]	(G) to node [auto] {$p_1$}	(T1);
					\draw[->,>=latex]	(G) to node [auto] {$p_1$}	(C1);

					\draw[->,>=latex]	(D) to node [auto] {$p_2$}	(T2);
					\draw[->,>=latex]	(D) to node [auto] {$p_2$}	(C1);

					\draw[->,>=latex]	(C1) to node [auto] {$p_1 \oplus p_2$}	(C2);

					\draw[->,>=latex]	(C2) to node [auto] {$p_1 \oplus p_2$}	(T1);
					\draw[->,>=latex]	(C2) to node [auto] {$p_1 \oplus p_2$}	(T2);
				\end{scope}
			\end{tikzpicture}
		}
		\label{Butterfly}
	}
	\caption{\label{NCIllustr} Network Coding Illustrations}
\end{figure}
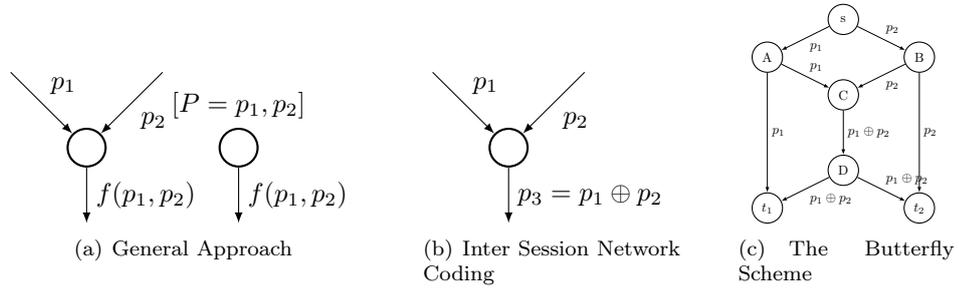
\subsubsection{Inter Flow}
\paragraph{Definition.}
Inter Flow (Inter Session) Network Coding relies on using the incoming packets to code the outgoing ones~\cite{Karl:2005:PAW:1076303,KKHR-CCC2005,citeulike:3701823,Ho:2008:NCI:1817176}.
Most of the time, the coding consists in a simple XOR between all the packets, as shown in \textbf{Fig.~\ref{InterFlowIllustr}}, where $p_1$ and $p_2$ are the incoming packets, and $p_3$ is the outgoing packet formed by the XOR between the two previous ones.

In order to illustrate the basics, we will see the butterfly scheme (\textbf{Fig.~\ref{Butterfly}}).
If a source $s$ has to send two packets $p_1$ and $p_2$ to both destinations $t_1$ and $t_2$, it can use Inter Flow Network Coding as follow:

\begin{itemize}
	\item $s$ sends $p_1$ to $A$, and $p_2$ to $B$.
	\item $A$ forwards $p_1$ to $C$, but when sending this message $t_1$ heard the transmission, and received $p_1$.
	\item In the same time, $B$ forwards $p_2$ to $C$, but $t_2$ also heard the transmission and received $p_2$.
	\item $C$ has to transmit $p_1$ and $p_2$.
	\item As the two packets have to reach the same destinations, it decides to transmit to $D$ a single packet corresponding to $p_1 \oplus p_2$.
	\item $D$ forwards the XORed packet to both $t_1$ and $t_2$.
	\item $t_1$ obtains $p_2$ by XORing the received packet with $p_1$ that it received earlier.
	\item $t_2$ does the exact same with $p_2$.
\end{itemize}

\paragraph{Benefits and limits.} 
\label{par:Advantages and limits}
As a summary, Inter Flow Network Coding is relevant because:

\begin{itemize}
	\item XOR is a trivial operation to code and decode.
	\item It may save bandwidth by sending one XORed packet instead of multiple ones.
\end{itemize}

But it has some constraints:

\begin{itemize}
	\item Nodes have to know what the other nodes heard.
\end{itemize}

Moreover there is a lack in Inter Flow Network Coding literature.

\subsubsection{Intra Flow}
\paragraph{Definition.}
In packets networks, Intra Flow (Intra Session) Network Coding consists in dividing a message (a data packet) into multiples submessages of the same size, and then creating a linear dependency between them before transmitting~\cite{Ho:2008:NCI:1817176}.

When the sink receives enough packets, it can recreate the initial message, by resolving the linear system created by the linearly dependent subpackets.

\paragraph{Advantages and limits.}
Intra Flow Network Coding could be a great tool in networks because of multiple assets:
\begin{itemize}
	\item It limits replication. In an optimal way, a packet is never resent.
	\item It creates redundancy. The number of coded packets can be greater than the number of divided packets, which means that information is created, but never duplicated.
	\item It improves the global reliability of the network protocol. If more coded packets are generated, if a certain amount of them are lost in the network, the sink can still decode the message if it receives enough coded packets.
\end{itemize}
On the other hand, Intra Flow Network Coding has some disadvantages:
\begin{itemize}
	\item It needs some computations on the source node and the sink, and sometimes on the intermediate nodes.
	\item It may lead to extra latency.
\end{itemize}

\paragraph{Random Linear Network Coding Spotlight.}
In order to explain the mechanisms used in Intra Flow Network Coding we will see how the Random Linear Network Coding works step by step~\cite{Ho:2008:NCI:1817176}.

\textbf{Step 1: The packet subdivision.}
As we have seen before, Intra Session Network Coding relies on the division of a data packet into a pre-defined number of same sized packets.
Here we consider a message as a chain of bits.

\begin{figure}
	\centering
	\includegraphics[width=200pt]{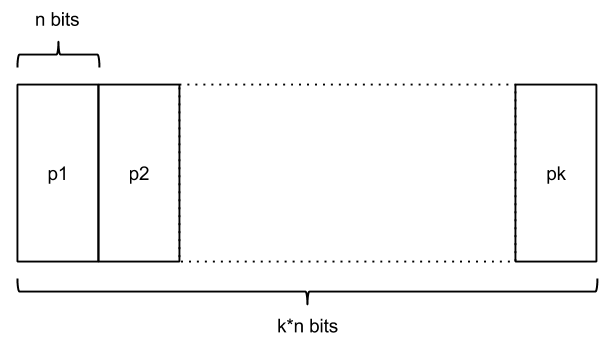}
	\caption{\label{Step1Illustr} Subdivision Illustration}
\end{figure}

The initial source node has to split a data packet into $k$ packets $p_1, p_2, ..., p_k$ of $n$ bits as illustrated in \textbf{Fig.~\ref{Step1Illustr}}.
This implies that the original message has to be a multiple of $n$.
Usually we take $n=8$ because it means that the coefficients are chosen in a Galois Field of size $2^8$.
This field allow the created packets to be linearly independent with a probability $P=0.996$, and every coefficient have the size of a $byte$, which represent a good compromise~\cite{NCMR}.

\textbf{Step 2: The coding coefficient choice.}
For each packet $p_i$, the node have to randomly choose $k' \geq k$ coefficients $<c^1_{p_i}, c^2_{p_i}, ..., c^{k'}_{p_i}>$ from a Galois Field $GF(2^8)$ to form the coefficients vectors.
In order to have good performances, the coefficients are picked in a precomputed Galois Field table.
Galois Field arithmetic is explained in \textbf{Appendix.~\ref{GFSec}}.

\textbf{Step 3: The coding.}
We put the previous vectors in a $k*k'$ matrix in order to obtain the coefficients matrix as below.

\begin{center}
	\begin{tabular}{c c}
		 & $k$ \\
		 & \\
		$\begin{matrix}\\k'\geq k\\\\\end{matrix}$ & $\begin{pmatrix}c_1^1 & c_2^1 & \cdots & c_k^1\\c_1^2 & c_2^2 & \cdots & c_k^2\\\vdots & \vdots & \ddots & \vdots\\c_1^{k'} & c_2^{k'} & \cdots & c_k^{k'}\end{pmatrix}$ \\
		 & \\
		 & $\begin{matrix}p_1 & p_2 & \cdots & p_k\end{matrix}$
	\end{tabular}
\end{center}

Then we create the encoded data $Y_j$ using the formula $Y_j=\sum_{i=1}^kc_i^jp_i$, with $j=1, ..., k'$.

\[
	\begin{pmatrix}
		c_1^1 & c_2^1 & \cdots & c_k^1 \\
		c_1^2 & c_2^2 & \cdots & c_k^2 \\
		\vdots & \vdots & \ddots & \vdots \\
		c_1^{k'} & c_2^{k'} & \cdots & c_k^{k'}
	\end{pmatrix}
	\rightarrow
	\begin{bmatrix}
		Y_1 = \sum_{i=1}^kc_i^1p_i \\
		Y_2 = \sum_{i=1}^kc_i^2p_i \\
		\vdots \\
		Y_{k'} = \sum_{i=1}^kc_i^{k'}p_i
	\end{bmatrix}
\]

\begin{figure}
	\centering
	\includegraphics[width=200pt]{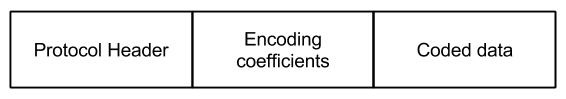}
	\caption{\label{Step4Illustr} Encoded Packet}
\end{figure}

\textbf{Step 4: The dissemination.}
Each $Y_j$ encoded data is then encapsulated with its coefficients vector $<c_1^j,c_2^j,...,c_k^j>$ in a packet (\textbf{Fig.~\ref{Step4Illustr}}) to be broadcast in the network, using any routing protocol.

Moreover, because we are in a Galois Field, each intermediate node receiving $b$ encoded packets $Y_1, Y_2,...,Y_b$ with their $c_1^i,c_2^i,...,c_k^i$ ($i=1,..,b$), can pick new encoding coefficients from $GF(2^8)$ and can create new packets as seen before.

\textbf{Step 5: The decoding.}
Whenever the sink receives $m$ packets, it puts the received coefficients vectors in a matrix.
If these coefficients satisfy the full rank matrix~\cite{DBLP:journals/jnca/YangZSY10}, which means that they are all linearly independent, the sink can retrieve the original subdivided messages by following this formula:

\[
	\begin{bmatrix}
		p_1 \\
		p_2 \\
		\vdots \\
		p_m
	\end{bmatrix}
	=
	\begin{pmatrix}
		c_1^1 & c_2^1 & \cdots & c_k^1 \\
		c_1^2 & c_2^2 & \cdots & c_k^2 \\
		\vdots & \vdots & \ddots & \vdots \\
		c_1^{m} & c_2^{m} & \cdots & c_k^{m}
	\end{pmatrix}
	^{-1}
	\begin{bmatrix}
		Y_1 \\
		Y_2 \\
		\vdots \\
		Y_m
	\end{bmatrix}
\]

We have to notice that the vector $<p_1, p_2, ..., p_m>$ is equal to the original one ($<p_1, p_2, ..., p_k>$), and is ordered no matter the $Y_i$ reception order (i.e. there is no particular order for the $Y_i$ vector's elements in order to decode).

The original data is then obtained by assembling the $p_i$ together.

\subsection{Routing Protocols using Network Coding}
\subsubsection{Routing Protocols that use Intra Flow Network Coding}
\paragraph{NC-RMR.}
NC-RMR stands for Network Coding Reliable braided and disjoint Multipath Routing~\cite{DBLP:journals/jnca/YangZSY10}.
In order to work, this routing protocol relies on Random Linear Network Coding to ensure the coding, and ReInForM for the number of paths computation and the next hop node selection.

\begin{figure}
	\centering
	\subfigure[NC-RMR Braided Multipath]{
		\scalebox{0.6}{
		\begin{tikzpicture}
			\begin{scope}
				\node[draw,circle,thick,minimum size=8mm]	(S) at (0,0)	{s};
				\node[draw,circle,thick,minimum size=8mm]	(11) at (2,2)	{};
				\node[draw,circle,thick,minimum size=8mm]	(12) at (2,0)	{};
				\node[draw,circle,thick,minimum size=8mm]	(13) at (2,-2)	{};
				\node[draw,circle,thick,minimum size=8mm]	(21) at (6,2)	{};
				\node[draw,circle,thick,minimum size=8mm]	(22) at (6,0)	{};
				\node[draw,circle,thick,minimum size=8mm]	(23) at (6,-2)	{};
				\node[draw,circle,thick,minimum size=8mm]	(T) at (8,0)	{t};

				\draw[->,>=latex]	(S) to node [auto] {$Y_1$}	(11);
				\draw[->,>=latex]	(S) to node [auto] {$Y_2$}	(12);
				\draw[->,>=latex]	(S) to node [auto] {$Y_3$}	(13);

				\draw[dashed,->,>=latex]	(11) to node [auto] {$Y_i^n$}	(21);
				\draw[dashed,->,>=latex]	(11) to node [auto] {}	(22);
				\draw[dashed,->,>=latex]	(11) to node [auto] {}	(23);
				\draw[dashed,->,>=latex]	(12) to node [auto] {}	(21);
				\draw[dashed,->,>=latex]	(12) to node [auto] {}	(22);
				\draw[dashed,->,>=latex]	(12) to node [auto] {}	(23);
				\draw[dashed,->,>=latex]	(13) to node [auto] {}	(21);
				\draw[dashed,->,>=latex]	(13) to node [auto] {}	(22);
				\draw[dashed,->,>=latex]	(13) to node [auto] {}	(23);

				\draw[->,>=latex]	(21) to node [auto] {$Y_1^n$}	(T);
				\draw[->,>=latex]	(22) to node [auto] {$Y_2^n$}	(T);
				\draw[->,>=latex]	(23) to node [auto] {$Y_3^n$}	(T);
			\end{scope}
		\end{tikzpicture}
		}
		\label{NCRMR}
	}
	\hspace{15pt}
	\subfigure[GBR-NC Workflow]{
		\scalebox{0.6}{
		\begin{tikzpicture}
			\begin{scope}
				\node[draw,circle,thick,minimum size=8mm]	(R1) at (0,0)	{$R_1$};
				\node[draw,circle,thick,minimum size=8mm]	(R2) at (4,0)	{$R_2$};
	
				\draw[->,bend left,>=latex]			(R1)	to node [auto] {$1$}	(R2);
				\draw[->,loop right,>=latex]		(R2)	to node [auto] {$2$}	(R2);
				\draw[->,>=latex]					(R2)	to node [above] {$3$}	(R1);
				\draw[->,loop left,>=latex]			(R1)	to node [auto] {$4$}	(R1);
				\draw[->,bend right,>=latex]		(R1)	to node [auto] {$5$}	(R2);
				\draw[->,out=300,in=240,>=latex]	(R1)	to node [auto] {$5$}	(R2);
			\end{scope}
		\end{tikzpicture}
		}
		\label{GBRNC}
	}
	\caption{\label{Both} Routing Protocols with Network Coding}
\end{figure}
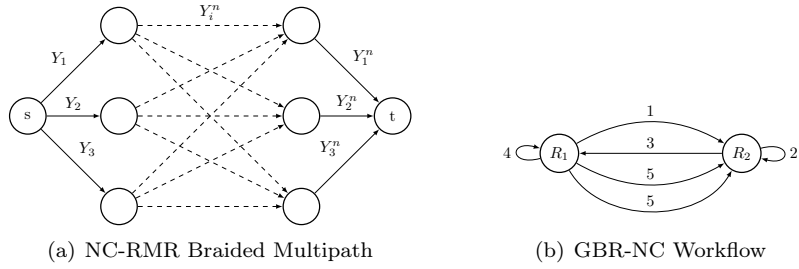

NC-RMR provides several features:

\begin{description}
	\item [Multipath on coded packets.] It is the major improvement over ReInForM, because there is no longer identical packets on the network.
		Each path will contain different packets created by the Random Linear Network Coding, which leads to redundancy reduction.
	\item [Hop by hop braided multipath.] In addition to the fact that a message is encoded at the source, every time a node receives a packet $Y_i$, it will generate $p$ new packets using Random Linear Network Coding.
		Then new paths are computed for each created packet, which leads to a braided multipath routing as in \textbf{Fig.~\ref{NCRMR}}.
		The message is then reconstructed as in standard Random Linear Network Coding.
\end{description}

\paragraph{GBR-NC.}
This protocol uses Random Linear Network Coding over Gradient Based Routing.
This algorithm adds the concept of Negative \textit{ACK} (\textit{NACK})~\cite{Miao2012990}.

Each node has a \textit{sending out number} initialised to $1$.
The \textit{sending out number} is a sort of redundancy needed in order to have good delivery ratio.
As illustrated in \textbf{Fig.~\ref{GBRNC}}, this routing protocol works as follows:

\begin{enumerate}
	\item A node receives a new encoded packet.
	\item The node tries to decode the message during a period $t$. During this period, it will try to use the other incoming coded message to decode.
	\item If the node did not manage to decode the packet, it will send a \textit{NACK} to it previous node.
	\item When a node receives a \textit{NACK}, it computes a value:
		$$R_{NACK}=\frac{number\ of\ received\ NACK}{number\ of\ sended\ messages}$$
		If $R_{NACK} \leq 0.05$, it increases its \textit{sending out number}.
	\item The node sends \textit{sending out number} packets.
\end{enumerate}

Every time a node has to send a packet over the network, it will send it a \textit{sending out number} of times.

\paragraph{S-NC.}
S-NC~\cite{DBLP:conf/mswim/Enayati-NoabadiYM12} is a routing protocol adding Network Coding (Inter Session) on a 802.11e like routing algorithm.

In fact it separates the nodes into three different categories:

\begin{itemize}
	\item A Class. The node is close to the sink, it has some packets in its queue and its battery is half filled.
	\item B Class. The node is close to the sink and it has some packets in its queue.
	\item C Class. The node is close to the sink.
\end{itemize}

Priority has been added to these classes, and A Class $>$ B Class $>$ C Class.

In order to route a packet to its next hop, this algorithm relies on the RTS/CTS principle.
In fact, whenever a node has to transmit a message, it will make a Request To Send.
Considering the priority classes, the node with the best properties is more likely to send a CTS back.
This mechanism allows the network to naturally load balance the energy consumption, while trying to have the shortest possible route.

\paragraph{NCMR.}
Network Coding Multipath Routing~\cite{NCMR} relies on Random Linear Network Coding in order to encode the packets.
In order to work, the sink first establishes the different routes by sending some RDP packets.
Then, whenever a node has to send a message, it computes the number $N$ of paths needed to achieve the delivery.
$N$ increases when the Bit Error Rate $BER$ and the number of hops $H$ needed to reach the sink increase.
Then it sends $N$ to the next node in its route table, the next node will then select $N-1$ backup nodes, and forward $N$.
The source will then code the message using Random Linear Network Coding, and send the generated packets over the network.
Whenever a node receives a coded message, it generates a local encoding matrix with coefficients randomly taken from $GF(2^8)$, and recodes the message before forwarding them.

In order to ensure the routes are well maintained, the sink periodically sends RMP.

\subsubsection{Routing Protocols that use Inter Flow Network Coding}

\paragraph{COPE Spotlight.}
COPE is a network architecture for wireless mesh networks~\cite{citeulike:3701823}, which relies on two fundamentals points:
\begin{itemize}
	\item The broadcast nature of the wireless channel. It means that we're not considering it as a point to point link.
	\item COPE relies on unicast traffic.
\end{itemize}

\begin{figure}
	\centering
	\subfigure[Without COPE]{
		\begin{tikzpicture}
			\node[draw,circle,thick,minimum size=8mm] (A) at (-2,0) {A};
			\node[draw,circle,thick,minimum size=8mm] (B) at (2,0) {B};
			\node[draw,circle,thick,minimum size=8mm] (R) at (0,0) {};

			\draw[->,>=latex] (-1.6,0.2) to node [auto] {1 : $p_1$} (-0.4,0.2);
			\draw[->,>=latex] (-0.4,-0.2) to node [auto] {4 : $p_2$} (-1.6,-0.2);
			\draw[->,>=latex] (1.6,-0.2) to node [auto] {2 : $p_2$} (0.4,-0.2);
			\draw[->,>=latex] (0.4,0.2) to node [auto] {3 : $p_1$} (1.6,0.2);
		\end{tikzpicture}
		\label{COPEIllustrWithout}
	}
	\hspace{15pt}
	\subfigure[With COPE]{
		\begin{tikzpicture}
			\node[draw,circle,thick,minimum size=8mm] (A) at (-2,0) {A};
			\node[draw,circle,thick,minimum size=8mm] (B) at (2,0) {B};
			\node[draw,circle,thick,minimum size=8mm] (R) at (0,0) {};

			\draw[->,>=latex] (A) to node [auto] {1 : $p_1$} (R);
			\draw[->,>=latex] (B) to node [auto] {2 : $p_2$} (R);
			\draw[->,>=latex] (R) to node [auto] {3 : $p_1 \oplus p_2$} (0,-2);
		\end{tikzpicture}
		\label{COPEIllustrWith}
	}
	\caption{\label{COPEIllustr} COPE Benefits}
\end{figure}
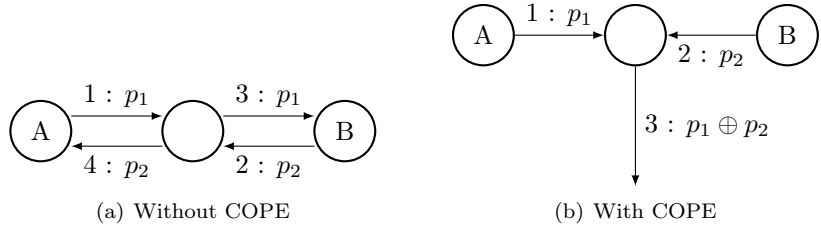

\textbf{Fig.~\ref{COPEIllustr}} shows the main benefits of COPE.
In a standard scenario (\textbf{Fig.~\ref{COPEIllustrWithout}}) where a node $A$ has to send a message $p_1$ to node $B$, and node $B$ has to send a message $p_2$ to node $A$, $A$ and $B$ will send their messages to an intermediate node/relay, which will then send $p_1$ to $B$ and then send $p_2$ to $A$.
In order to achieve the deliveries, $4$ transmissions have been used.

On the other hand (\textbf{Fig.~\ref{COPEIllustrWith}}), with COPE, when the intermediate node has to send the messages it will "broadcast" only one message $p_1 \oplus p_2$.
In order to do that, we're considering that $A$ and $B$ are overhearing on the radio channel, and that the intermediate node is able to know that both $A$ and $B$ have enough information in their buffer to decode.

\begin{figure}
	\centering
	\subfigure[Bad Coding: P1 $\oplus$ P2, A can't decode P1]{
		\label{BadCoding}
		\scalebox{0.6}{
			\includegraphics[width=150pt]{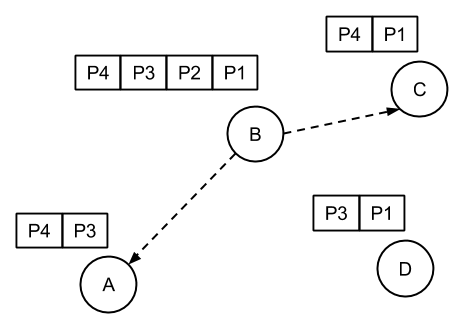}
		}
	}
	\subfigure[Good Coding: P1 $\oplus$ P3, A and C can decode]{
		\label{GoodCoding}
		\scalebox{0.6}{
			\includegraphics[width=150pt]{COPEOCBad.png}
		}
	}
	\hspace{5pt}
	\subfigure[Best Coding: P1 $\oplus$ P3 $\oplus$ P4, A, C and D can decode]{
		\label{BestCoding}
		\scalebox{0.6}{
			\includegraphics[width=150pt]{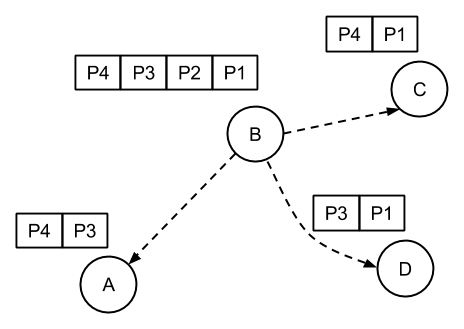}
		}
	}
	\caption{\label{OppCoding} Opportunistique Coding. Buffers are: A(P3, P4), B(P1, P2, P3, P4), C(P1, P4) and D(P1, P3). Destinations are: P1$\rightarrow$A, P2$\rightarrow$C, P3$\rightarrow$C, P4$\rightarrow$D}
\end{figure}

COPE acts as a coding layer between the network layer (IP in this case) and the MAC layer.
In order to do this, it relies on three main techniques:
\begin{description}
	\item [Opportunistic Listening.] COPE's nodes benefit from the wireless medium's properties, by snooping on all communications and storing the overheard packets.
		Each packet possesses a particular header, see \textbf{Appendix.~\ref{COPEHeader}}, containing the XORed data, the packets used to XOR IDs, a \textit{Reception Report} containing all the packets the sender has in its buffer, and a bunch of \textit{ACK}.
	\item [Opportunistic Coding.] The objective is to \textit{maximize the number of original packets delivered in a single transmission, while ensuring that each intended next hop has enough information to decode its native packet}, as illustrated in the scenario of \textbf{Fig.~\ref{OppCoding}}.
		To ensure that all next nodes of an encoded packet can decode their corresponding packets, COPE establishes a rule:
		\begin{quotation}
			To transmit $n$ packets $p_1,...,p_n$, to $n$ next hops $r_1,...,r_n$, a node can XOR the $n$ packets together only if each next hop $r_i$ has all $n-1$ packets $p_j$ for $j \neq i$.
		\end{quotation}
	\item [Opportunistic Routing.] The idea is to avoid sending a packet to a node that already get it when overhearing.
		COPE uses a guessing system based on geographical, or ETX metric to do so.
\end{description}

\section{The use of network coding in Resiliency}
\label{Resiliency}

\subsection{Problem Statement}
We propose here to see the effects of Network Coding on resiliency, in presence of malicious nodes.

Malicious nodes are nodes that have been compromised by an adversary in order to disrupt the routing process.
A compromised node attempts several attacks on the network like droping packets instead of retransmitting them.

Wireless Sensor Networks should be resilient to this kind of behavior.
Resiliency is the ability for an entity to absorb negative behavior~\cite{erdene2011enhancing}.
More specificaly, for a Wireless Sensor Network, resiliency characterizes the ability of the network in handling malicious nodes.

\subsection{Back to the Resiliency Metric.}
\subsubsection{Definition.}
In order to study the impact of Network Coding in Wireless Sensor Networks, our work evaluation is based on the resiliency~\cite{5478822,6423640,erdeneochir:tel-00862710}.
In fact, it is difficult to have a metric in order to characterize wireless network.
Delivery Ratio is not enough to define the quality of a Routing Protocol, and the author would like to have a metric taking multiple parameters into account.

That is why the author uses a multi dimensional graphical metric, letting aggregate much more information.
Basically, for a number of parameters $n \geq 3$, and different percentages $k$ of compromised nodes, the metric will work as follows.
First, the different parameters' values $p_i(k)$ ($i=0,...,n$) are normalised in order to be "compared", following this formula:

$$p_i(k) = \frac{p_i(k)}{\max(\forall k, p_i(k))}$$

If the goal is to compare $m$ different protocols, the formula slightly differ:

$$p_{i,j}(k) = \frac{p_{i,j}(k)}{\max(\forall k,j, p_{i,j}(k))},\ j=1,...,m$$

\begin{figure}
	\centering
	\subfigure[$n = 3$]{
%
		\scalebox{0.3}{
			\includegraphics{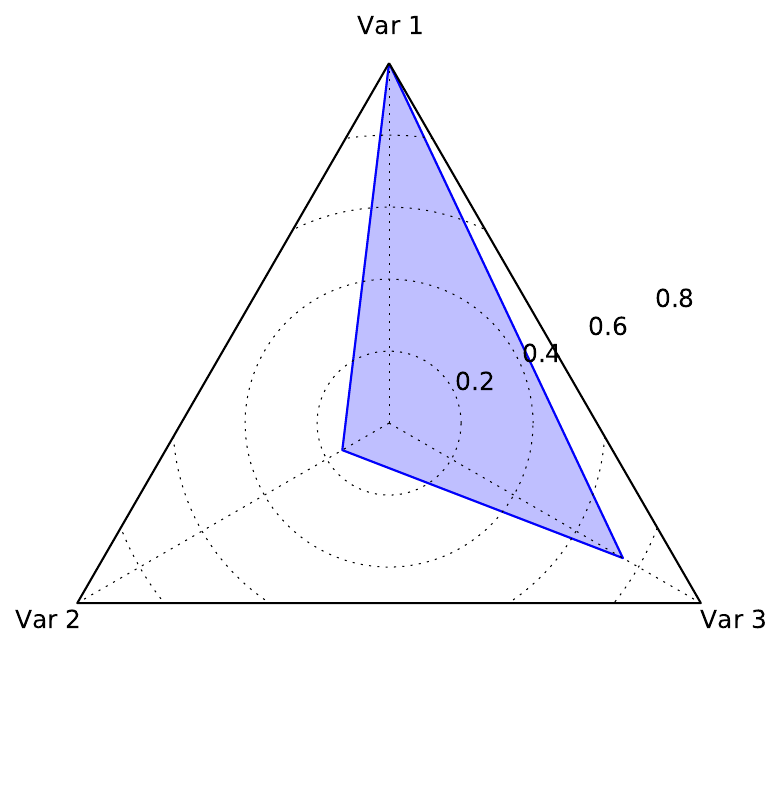}
		}
	}
	\hspace{15pt}
	\subfigure[$n = 5$]{
%
		\scalebox{0.3}{
			\includegraphics{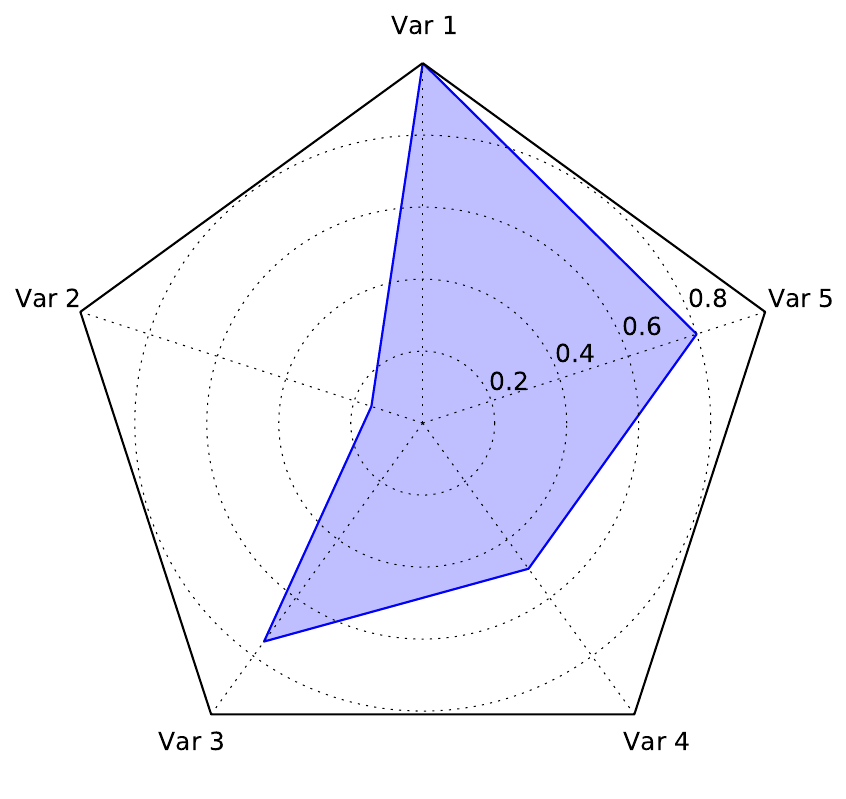}
		}
	}
	\caption{\label{GraphicalIllustr} Graphical representations with different parameters number}
\end{figure}

Then the different $p_i$ could be represented on an $n$ edged polygon as illustrated in \textbf{Fig.~\ref{GraphicalIllustr}}.

Then, in order to aggregate the information, the author proposes to compute the area made by the polygon, given the resiliency metric:

$$R_i(k)=(\sum_{j=1}^{n-1}(p_{i,j}(k)p_{i, j+1}(k))+p_{i,n}p_{i,1})\frac{1}{2}\sin(\frac{2\pi}{n})$$

A routing protocol is defined as resilient when the previous resiliency metric does not decrease much while $k$ increases.

\subsubsection{Resiliency's Parameters.} 
\label{ssub:Parameters}
The way we previously presented Resiliency was generic.
In fact we did not discuss the parameters we have to take in order to match a Wireless Sensor Network context.
That is why we decided to based our work on the five following parameters:
\begin{description}
	\item[Average Delivery Ratio (ADR).] $ADR = \frac{Number\ of\ packets\ received\ by\ the\ sink}{Number\ of\ packets\ sended\ by\ nodes}$.
		In terms of data delivery, a good delivery ratio is what Wireless Sensor Networks aim.
	\item[Delay Efficiency (DE).] It is the average end to end delay.
		This represents the time taken by a packet from sending to reception/decoding.
		Because of the nature of this parameter, we will consider the less delay, the better it is.
	\item[Energy Efficiency (EE).] It quantify the energy consumed by nodes in order to make the routing work.
		It takes \textit{Control Packets} and \textit{Data Packets} into account.
		For the same reasons than the Delay Efficiency, we will consider the less energy used as the better.
		By the way, as the overal energy consumed decreases in presence of malicious nodes, due to packets dropping, this parameter is computed as follow:
		$$p_{i,j}(k) = \frac{p_{i,j}(k)}{\max(p_{i,j}(k))},\ j=1,...,m$$
	\item[Average Throughput (AT).] It is the average amount of data flows the sink receives per source per unit of time.
	\item[Delivery Fairness (DF).] It is the deviation of each node's delivery ratio from the average.
		It may indicate the overall delivery distribution.
		As the compared routing protocol are similarly based, this parameter does not vary much.
\end{description}

\subsection{Network Coding for Resiliency}
In order to show the benefits of Network Coding in terms of Resiliency, we designed an Intra Session Network Coding Random Gradient Based Routing (RS-GBR-NC) protocol.
We base our work on Random Gradient Base Routing (RS-GBR) because it has been shown that it is the better gradient based resilient routing protocol~\cite{6423640}.

\subsubsection{Random Gradient Based Routing (RS-GBR).}
In order to improve global delivery ratio, this protocol proposes to avoid deterministic routing, by introducing some random choices in the next hop selection.
In fact, before forwarding, a node randomly chooses a next hop, with a probability varying with the distance to the sink of the possible next hop~\cite{6423640}.
More specifically, a node divides its next hop node possibilities into two groups:
\begin{itemize}
	\item Nodes with $next\ hop\ gradient = node's\ gradient - 1$.
		Then, the node will chooses a random node in this group, with a probability $P(next\ hop) = 0.8$.
	\item Nodes with $next\ hop\ gradient = node's\ gradient$.
		Then, the node will chooses a random node in this group, with a probability $P(next\ hop) = 0.2$.
\end{itemize}
This implies longer routes when the next hop node has the same gradient, but it allows greater routes possibilities.

This way, routes are never the same, and malicious nodes are not as powerful as if they were on a static route.

\subsubsection{Random Gradient Based Routing with Replication (RM-GBR).} 
\label{ssub:Random Gradient Based Routing with Replication (RM-GBR).}
In order to solve the problems addressed when a malicious node is on the path of a random route, the protocol may introduce some redundancy.
In fact a node can replicate a packet a chosen number of times, and can send it the same number of times over different randomly chosen paths.

\subsubsection{Random Gradient Based Routing with Network Coding (RS-GBR-NC).} 
\label{par:Random Gradient Based Routing}
In this alternative protocol, we decided to only code the packet at the source.
The number of generated packets is statically fixed.
It means that whenever a node has to send a message, it generates a fixed number of encoded messages that will take different randomly chosen paths.
When the sink receives enough packets, it decodes the message, and drops any other related incoming packets.

In order to explain this protocol mechanism, we propose to illustrate it with an example.

We consider an original message, as a $32$ bits word.
As we want to use coefficients from $GF(2^8)$, we have to divide the previous message into four $8$ bits submessages.
Then we have to generate coded packets.
Each packet generated from the same original message, has to have an ID identifying the original message, the coded data, and the coding vector used to code the submessage.
In order to retrieve the original message, we need four coded packets at least.
Because of packet loss and malicious nodes, if we only generate four packets, we will observe bad delivery ratio because the loss of one packet implies the loss of the entire message.
That is why we decide to generate $16$ coding packet at the source.
Then, for each coded packet, a next hop is randomly chosen using RS-GBR method, and this until the packet reach the sink.
When the sink receives a packet, it checks if it possesses enough packets with the same original message ID, in order to retrieve the message.
If there is enough packets, the sink starts to decode the message using a Gaussian elimination.
Then it clears its buffer from the used coded packet, and drops all the incoming packets with this ID.
If there is not enough packets, the sink stores the received packet in order to decode it later.

With this protocol, we expect to have a better delivery ratio, because Random Linear Network Coding introduces some redundancy over the network, which implies that the lost packets will be less painful.
But because of the nature of Random Linear Network Coding, more packets will transit over the network.

\subsubsection{Random Gradient Based Routing with Network Coding and Acknowledgment (RS-GBR-NCACK).} 
\label{par:Random Gradient Based Routing ACK}
This protocol is a slight improvement of the previous one.
In this protocol we are not generating and sending a statical amount of coded packets.
Instead we introduce an acknowledgement mechanism allowing us to dynamically generate coded packets.

\begin{figure}
	\centering
	\scalebox{0.5}{
		\includegraphics{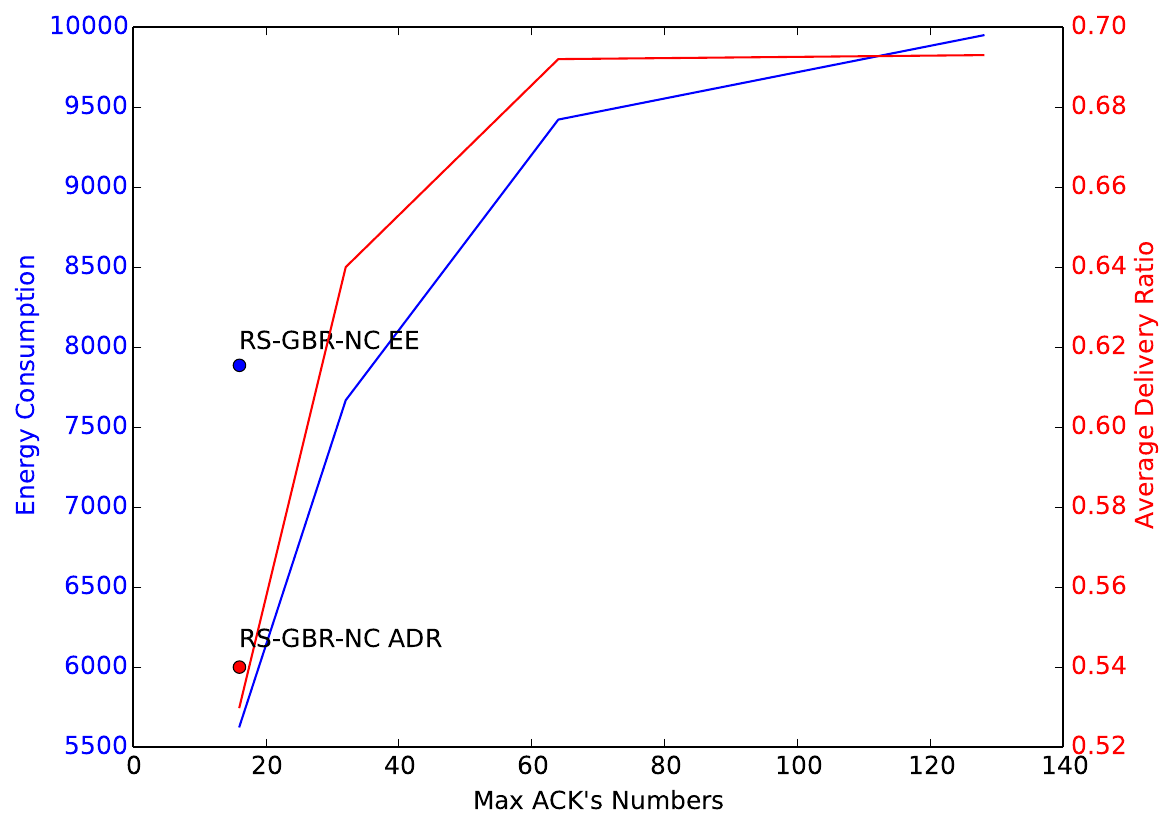}
	}
	\caption{\label{ACK} Average Delivery Ratio and Energy Consumption depending on the maximum number of coded packets generated while waiting for an ACK with 50\% malicious nodes}
\end{figure}

In fact as soon as the sink achieve a successful decoding, it sends an acknowledgement message to the sensor in order to tell it to stop its coded packet generation.
To achieve this goal we add a little delay between each coded packet generation in order to ensure that each coded packet posses enough time to reach the sink in order to avoid useless packets generations.
Moreover, some sensors cannot send packet to the sink because their neighborhood is only composed of  malicious nodes.
This lead to an infinite coded packet generation, and a greater energy consumption for this nodes while they are waiting for the sink acknowledgement.
In order to avoid this effect, we decide to cap the number of coded packet generated to 32.
As we can see in \textbf{Fig.~\ref{ACK}}, this maximum number allow the system to increase the overall average delivery ratio, while keeping the same energy consumption as the previous protocols.

\subsection{Results and Analysis}

\subsubsection{Simulation Parameters.}
In order to evaluate our work, we used WSNET simulator.
Simulations were performed 100 times to guarantee reasonable confidence interval.

Each simulation was made on a $100m * 100m$ square with $300$ nodes, following a poisson process distribution, and a single sink in the center of it.
Sensors had a $20m$ transmission range, and the average degree was $31$.

\subsubsection{Results.} 
\label{ssub:Results and Analysis}

The focus of our simulations is the comparaison of four routing protocols:

\begin{itemize}
	\item Random Gradient Based Routing (RS-GBR)
	\item Random Gradient Based Routing with Replication (RM-GBR)
	\item Random Gradient Based Routing with Network Coding (RS-GBR-NC)
	\item Random Gradient Based Routing with Network Coding and Acknowledgement (RS-GBR-NCACK)
\end{itemize}

\begin{figure}
	\centering
	\subfigure[RS-GBR]{
		\scalebox{0.25}{
			\includegraphics{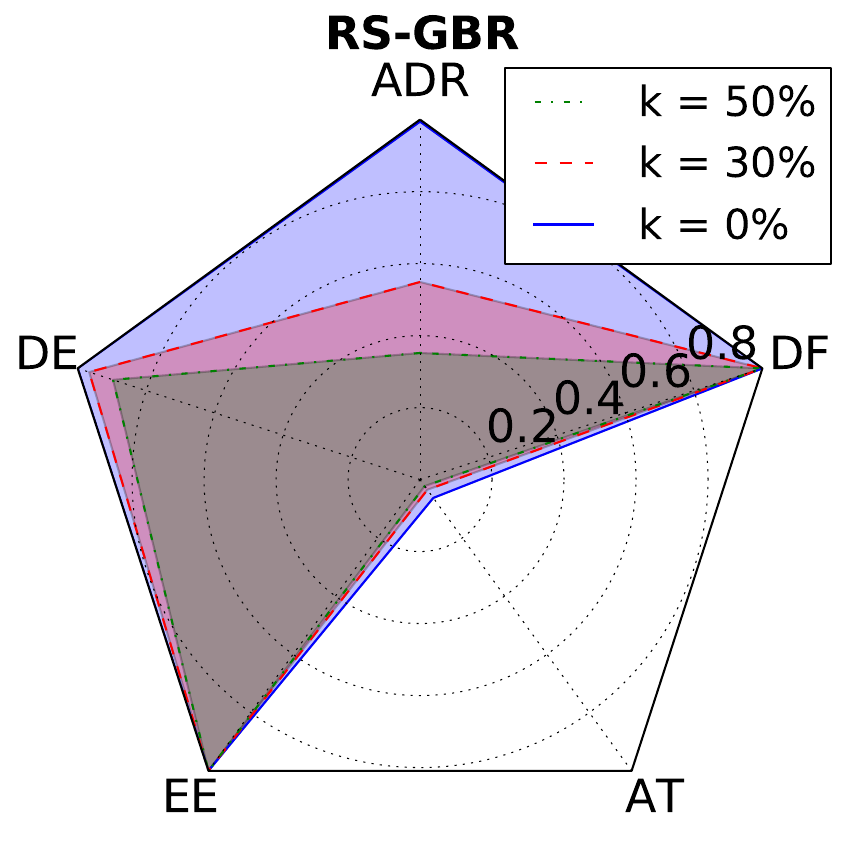}
		}
		\label{RSGBRKiviat}
	}
	\subfigure[RM-GBR]{
		\scalebox{0.25}{
			\includegraphics{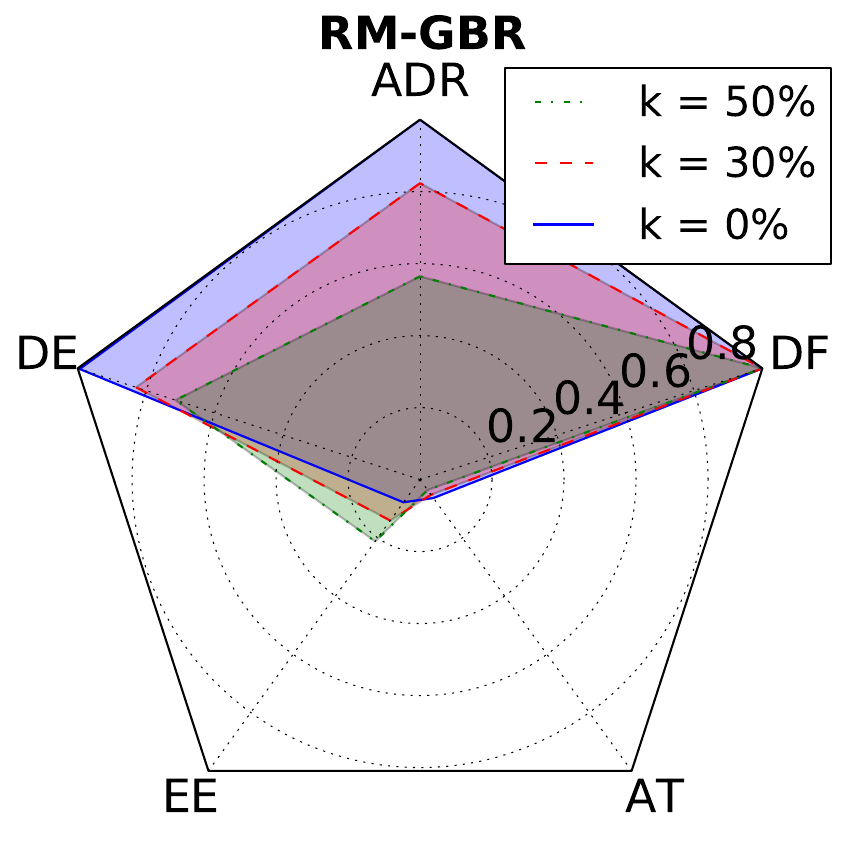}
		}
		\label{RMGBRKiviat}
	}
	\subfigure[RS-GBR-NC]{
		\scalebox{0.25}{
			\includegraphics{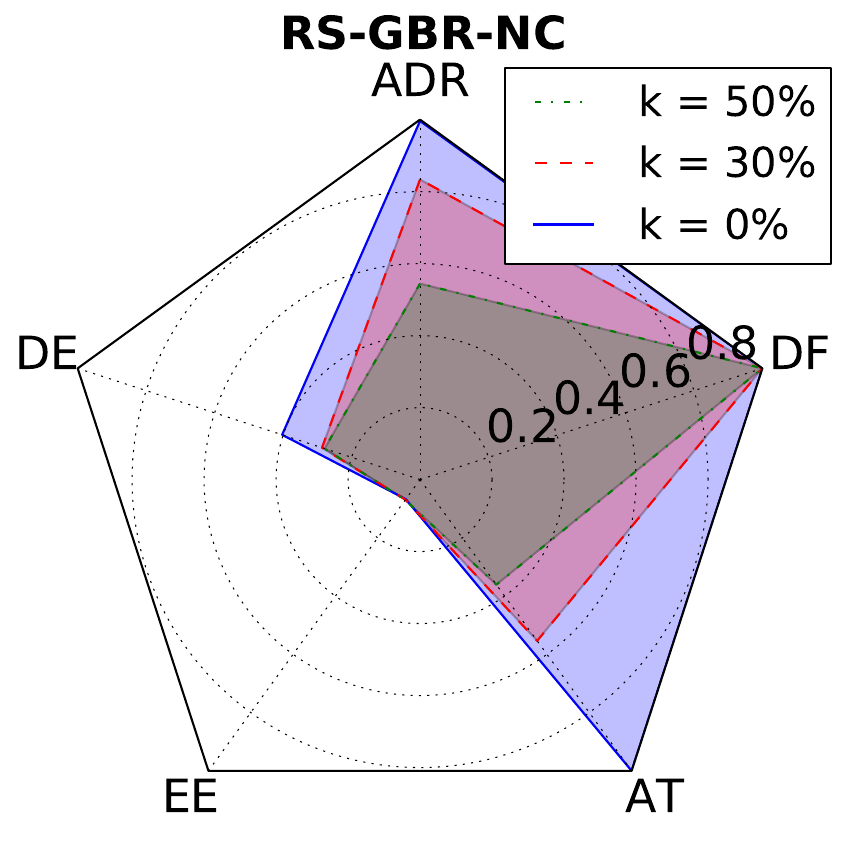}
		}
		\label{RSGBRNCKiviat}
	}
	\subfigure[RS-GBR-NCACK]{
		\scalebox{0.25}{
			\includegraphics{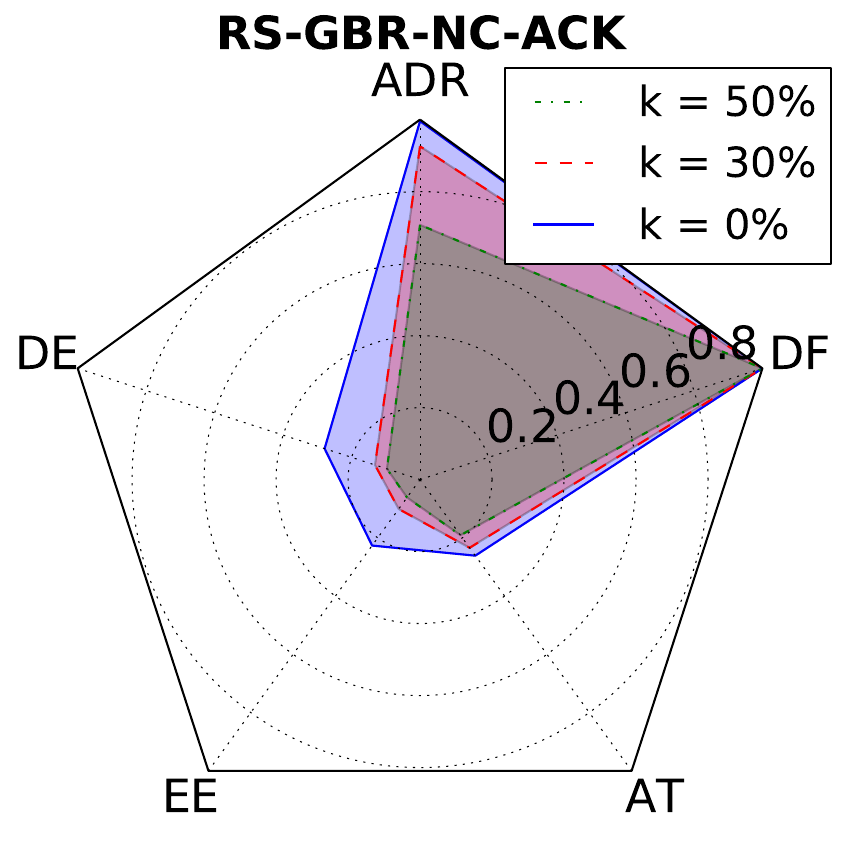}
		}
		\label{RSGBRNCACKKiviat}
	}
	\caption{\label{Kiviat} Resiliency Area depending on the percentage of Malicious Nodes}
\end{figure}

\paragraph{Graphical representation.} 
\label{par:Graphical representations analysis.}
Graphical representation of the different parameters are showed in \textbf{Fig.~\ref{Kiviat}}.
As we can see the different protocols exhibit different behaviors.

First, we observe that there is a gap between RS-GBR on one hand, and RM-GBR, RS-GBR-NC and RS-GBR-NCACK on the other hand, concerning the Energy Efficiency.
In fact, our energy model before normalization is one bit transmitted costs one, one bit received costs two.
As RM-GBR, RS-GBR-NC and RS-GBR-NCACK transmits more packets than RS-GBR in order to achieve a successful delivery, after the normalization, their Energy Efficiency is close to zero.
Concerning the acknowledgement improvement over RS-GBR-NC, we observe that it allows a better energy efficiency when there is few malicious nodes in comparaison to the version without acknowledgement.

Secondly, we observe the same behavior with the Average Throughput.
As RS-GBR-NC generates $16$ times more packets, and because we never saturate the network capacity, this protocol presents a huge throughput compared to RS-GBR, RM-GBR and RS-GBR-NCACK.
This leads to the fact that their metrics are close to zero, and that the throughput variations are not significant.

Finally, we observe quite the same thing concerning the Delay Efficiency, but the result is not as sharply contrasted as the previous ones, which let us observe the variations.
RS-GBR-NC is less efficient due to the fact that the decoding to the sink is time consuming, and that it has to wait until there is enough packets in order to decode.
Due to the delay introduction in RS-GBR-NCACK in order to mitigate the number of coded packets generated, this version is slightly below RS-GBR-NC in terms of Delay Efficiency.

\begin{figure}
	\centering
	\scalebox{0.5}{
		\includegraphics{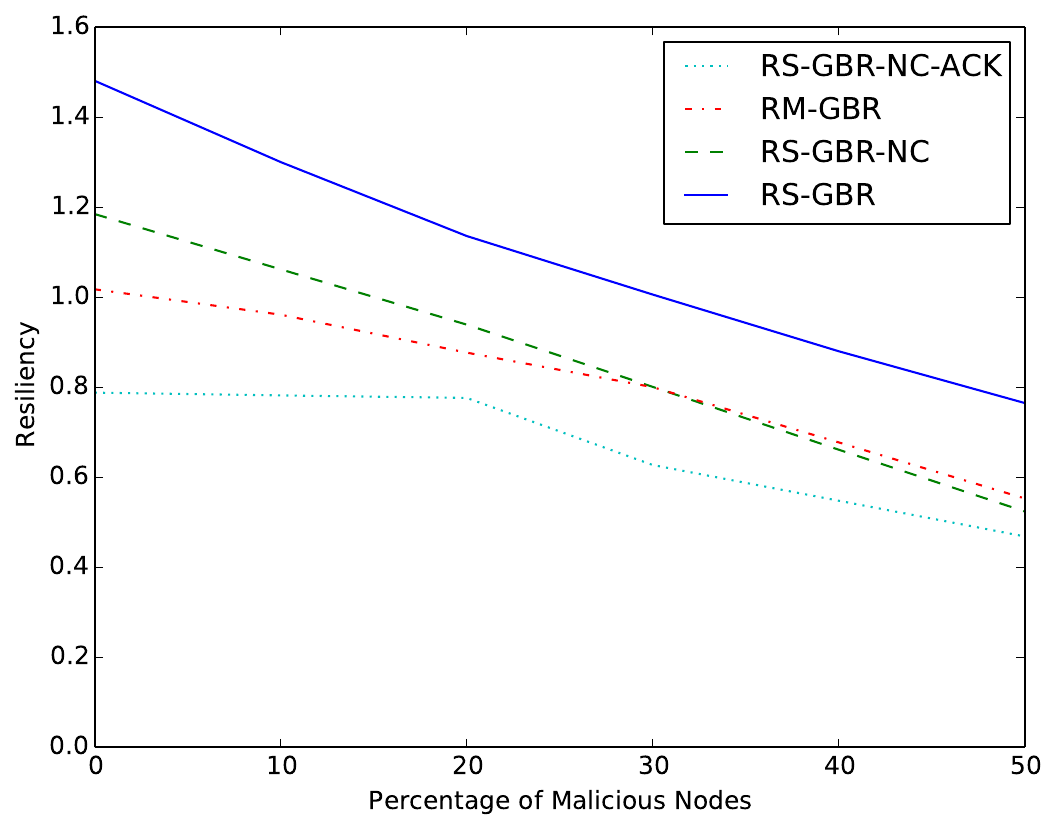}
	}
	\caption{\label{ResiliencyResult} Resiliency Evaluation in Presence of Malicious Nodes}
\end{figure}

\paragraph{Resiliency.}
The Resiliency evolution over the number of malicious nodes is showed in \textbf{Fig.~\ref{ResiliencyResult}}.

As we can see, RS-GBR has better Resiliency over RS-GBR-NC and RM-GBR.
This is mostly due to the fact that RS-GBR has greater Energy Efficiency than the others.
But the Resiliency evolution of RS-GBR, is not as good as the two others.

In fact, we see that RS-GBR's Resiliency falls quicker than the others when the malicious nodes increase from $0\%$  to $20\%$.
This means that this protocol is less resilient to the malicious nodes increase.

At the opposite, we observe that RM-GBR's Resiliency falls less quicker during the same malicious nodes increase, which means that it is more resilient in the first increases.
This is due to the fact that RM-GBR introduces redundancy, which mean that even if there is a malicious node on one path, a duplicated packet takes another path which could be malicious nodes free.
We observe very low variations on RS-GBR-NCACK's resiliency when there is few malicious nodes.
This flat curve means that RS-GBR-NCACK is very resilient when the number of malicious nodes begin to increase.
But this protocol has a low resiliency value because the flat evolution has a strong cost in termes of Delay Efficiency and Energy Efficiency.

Finally, RS-GBR-NC has a constant decrease, which means that it is resilient to the malicious nodes increase with the same intensity.

\subsubsection{Analysis.} 
\label{ssub:Analysis.}
Because of the nature of the resiliency computation, the order of the parameters matters.
In fact, they are acting as a weighting of each other.
As we use the same experimental protocol as~\cite{erdene2011enhancing}, Delay Efficiency weights the Average Delivery Ratio, and Energy Efficiency weights both Delay Efficiency and Average Throughput.
Because of the sharp contrasts concerning Delay Efficiency, Energy Efficiency and Average Throughput, these parameters order may not fit, and an other one could be more relevant.
Especially because we deal with two energy consuming protocols, which may lead to reconsidering the position of the Energy Efficiency.

As we deal with protocols using redundancy in order to achieve good deliveries, the Average Throughput may introduce a bias until we are not achieving the network capacity.
We may think of using a better parameter like "Goodput", which may characterize the quantity of useful data received by time unit.

\section{Conclusion and Further Works}
\label{Conclusion}

In this paper we have studied the impact of Network Coding in Wireless Sensor Networks. We add Network Coding to a resilient routing protocol, in order to quantify the influence of Network Coding in malicious environments.
Even if there is a bias due to the Energy Efficiency, we observe an improvement in Average Throughput and Average Delivery Ratio over RS-GBR.
Moreover, RS-GBR-NC has better Resiliency than RM-GBR.

This last study offers a lot of possible further works.
In a first time, it would be interesting to study the new Resiliency value when changing the parameters order to a better one.
It would be interesting to replace the Average Throughput by a "Goodput" parameter.

Then we envisage to replace the static number of generated packets, by the result of the distance to sink function.
We would like to introduce a system of \textit{ACK} too, in order to limits encoded generation when a good reception is done.
Moreover, it would be a great improvement, if we add Network Coding on each hop, in order to add some redundancy.

Finally, we would like to implement RS-GBR-NC on a real Wireless Sensor Network environment.


\bibliographystyle{plain}
\bibliography{biblio}

\appendix

\section{COPE's Header and Pseudo Broadcast} 
\label{COPEHeader}
\begin{figure}[ht!]
	\centering
	\includegraphics[width=200pt]{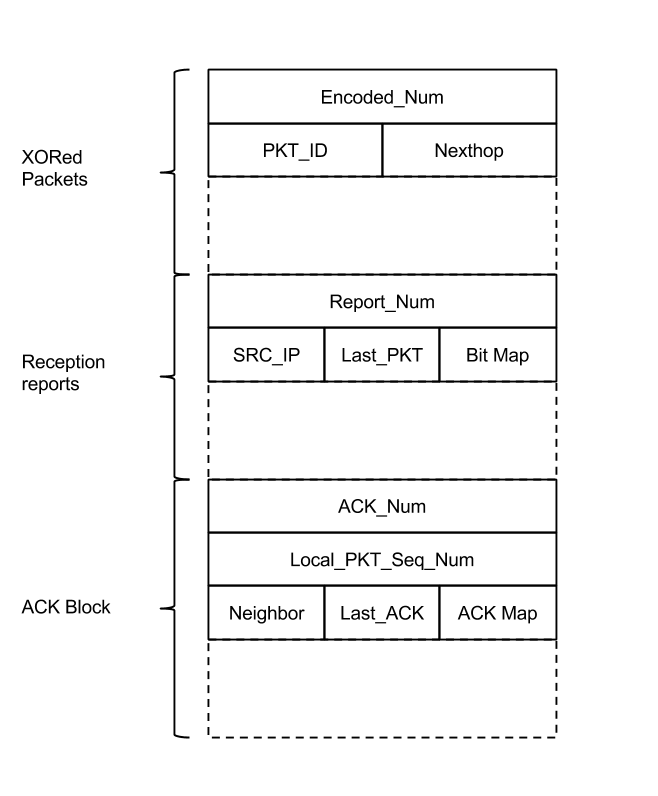}
	\caption{\label{COPEPacket} COPE Header}
\end{figure}

To fully understand how COPE~\cite{KKHR-CCC2005,4454988,citeulike:3701823} works, we have to highlight COPE's header (\textbf{Fig.~\ref{COPEPacket}}) and the pseudo broadcast.

COPE's header contains 3 defined blocks.
\begin{itemize}
	\item XORed Packets.
		$Encoded\_Num$ refers to the number of XORed packets the packet contain.
		It is followed by a list of size $Encoded\_Num$ containing a $PKT\_ID$ refering to the $ID$ of one of the packet used to XOR, and $Nexthop$ the destination of this packet.
	\item Reception Reports.
		$Report\_Num$ corresponds of the total size of the report.
		A list containing on each row the $SRC\_IP$, which is an overheard node's $IP$, $Last\_PKT$ which is the $ID$ of the last packet heard from the $SRC\_IP$, and a $Bit\ Map$ corresponding to the last packets from $SRC\_IP$ counting from $Last\_PKT$.
		For example, if $Last\_PKT=33$ and $Bit\ Map=10011$, it means that the node heard packets $33$, $32$, $31$ and $28$.
	\item ACK Block.
		Whenever a packet is sent, it contains some of the $ACK$ the node has to send.
		This mechanism limits the global $ACK$ overhead.
		$ACK\_Num$ stands for the number of $ACK$ entries.
		$Local\_PKT\_Seq\_Num$ is sort of the packet $ID$.
		It is the value of a counter that each node keeps for each of its neighbor, and increases when a message is sent to it.
		$Neighbor$ is the MAC Address of the neighbor the $ACK$ is addressed to.
		$Last\_ACK$ is a pointer to the $ID$ of the last received packet it wants to acknowledge.
		$ACK\ Map$ acts like $Bit\ Map$, but for the $ACK$.
\end{itemize}

Pseudo broadcast relies on the fact that despite packets are sent in a unicast way, every node overhears on the radio medium and checks if they are the destination.
Even if it is not the case, each node stores the data and $receptions\ reports$ contained in the packet.


\section{Galois Field} 
\label{GFSec}
This appendix aim at giving the minimum elements concerning Galois Field, in order to understand Random Linear Network Coding.

$GF(p^n)$ is a Finite Field containing elements from $0$ to $p^n-1$, and operations like addition, multiplication and their inverse.
These operations are defined in a way such that the result is always contained between $0$ and $p^n-1$ (closure of the Finite Field).

The previous operations have to satisfy some rules:

\begin{itemize}
	\item Commutativity
	\item Associativity
	\item Distributivity
	\item Every element $a$ has to have an additive inverse ($-a$) such that $a + (-a) = 0$
	\item Every element $a$ has to have a multiplicative inverse ($\frac{1}{a}$) such that $a * \frac{1}{a} = 1$
\end{itemize}

Galois Field's arithmetic is modular arithmetic, and it looks simpler to illustrate it with an example, the prime Galois Field $GF(2)$ (\textbf{Fig.~\ref{GF2}}).
The operations will be standard addition and multiplication modulo 2.
Elements in this field are $\{0, 1\}$.

\begin{figure}
	\centering
	\begin{tabular}{c c c}
		$0+0=0$ & \hspace{15pt} & $0*0=0$\\
		$0+1=1$ &     & $0*1=0$\\
		$1+1=0$ &     & $1*1=1$
	\end{tabular}
	\caption{\label{GF2} $GF(2)$ Operations}
\end{figure}

To ensure that we're defining a Field and not a Ring, in $GF(m)$, $m$ has to be a prime number.
In the other case, some elements in the Field will not have a multiplicative inverse.

But if $m$ is a prime power ($m=p^n$ with $p$ a prime number), we can act as we are working on a field, but we have to redefine how addition and multiplication work.
Moreover, we still make operation on elements contained in $0, ...,  p^n-1$, but the operations will be easier to understand if we represent each element as a polynomial.
In fact, every element $a \in {0, ..., p^n-1}$ is represented with $n$ base$-p$ digits $a_0, ..., a_{n-1}$:
$$a=a_0 + a_1p + a_2p^2 + ... + a_{n-1}p^{n-1}$$
Now we can consider $a$ as a vector made of these coefficients. For example a binary digit modulo $2^n$ is represented by a chain (vector) of $n$ bits.
In field with $p^n$ elements, the addition is just like a vector addition.
For example in $GF(4) = GF(2^2)$, the addition looks like in \textbf{Fig.~\ref{GF4}}.
Elements in $2^n$ Field are represented as $a=a_0 + a_12 + a_24 + ... + a_{n-1}2^{n-1}$ with $a_i = {0,1}$.
\begin{figure}
	\centering
	\begin{tabular}{c c c}
		$00+00=00$ & \hspace{15pt} & $01+10=11$\\
		$00+01=01$ & \hspace{15pt} & $01+11=10$\\
		$00+10=10$ & \hspace{15pt} & $10+10=00$\\
		$00+11=11$ & \hspace{15pt} & $10+11=01$\\
		$01+01=00$ & \hspace{15pt} & $11+11=00$
	\end{tabular}
	\caption{\label{GF4} $GF(4)$ Additions represented in binary}
\end{figure}
Note that in this field, addition looks like $XOR$ which could be considered as an addition without carries.

The multiplication of two elements in the Field consists in multiplying their polynomial representations, but still taking the modulo into account.
But as we can see, the result may sometimes exceed the range of the field.
That's why we will have to reduce the resulted polynomial, with a well chosen one (irreductible, and of order $n$, i.e. primitive) $P(X)$.
In fact every time a result exceeds the range (polynomial degre $\geq n$), we made a polynomial long division by $P(X)$ of it.

We are not going to explain how the particular polynomial is chosen, but we will consider that for $GF(2^8)$, it will be $P(X)=X^8+X^4+X^3+X^2+1$ (remember that as $p=2$, coefficient are ${0, 1}$.

Finally, as polynomial multiplication (and division) are resource consuming, it is easier to go in a logarithmic representation.
It rely on the fact that we can multiply two numbers $p$ and $q$ with the following operation $b^{\log_b{p}+\log_b{q}}$.
As we are in base $2$ we will use $2^{\log_2{p}+\log_2{q}}$, and we have to consider that the addition is still modulo $2$.
Moreover, in $GF(2^8)$, every element can be represented as a power of $2$. The value are obvious for the numbers in $[2^0, 2^7]$, but when the exponent is greater or equal to $8$, the value is greater than $255$, $2^8=256$.
When this happens, we have to reduce the value, which is equal to add the value with $P(X)$:
\[
	X^8 + X^8 + X^4 + X^3 + X^2 + 1 = 2X^8 + X^4 + X^3 + X^2 + 1 = X^4 + X^3 + X^2 + 1
\]
\[
	\Leftrightarrow 256 \oplus 285 = 29
\]
Then we can find a logical recursion: $2^9=2*2^8=58$, $2^{10} = 2*2^9$, ...
And every time the result exceeds $255$, we XOR it with $285$.
But as logarithmic representation is resource consuming too, we will pre compute this result in lookup tables, as in \textbf{Fig.~\ref{LOGTable}}.
\begin{figure}[ht!]
	\centering
	\begin{tabular}{l | r | c | l | r}
		\textbf{Exponent} & \textbf{Value} & \hspace{15pt} & \textbf{Value} & \textbf{Exponent}\\
		$0$ & $1$ & & & \\
		$1$ & $2$ & & $1$ & $0$\\
		$2$ & $4$ & & $2$ & $1$\\
		$3$ & $8$ & & $3$ & $25$\\
		$4$ & $16$ & & $4$ & $2$\\
		$5$ & $32$ & & $5$ & $50$\\
		$6$ & $64$ & & $6$ & $26$\\
		$7$ & $128$ & & $7$ & $198$\\
		$8$ & $29$ & & $8$ & $3$\\
		$9$ & $58$ & & $9$ & $223$\\
		$\vdots$ & $\vdots$ & & $\vdots$ & $\vdots$\\
		$255$ & $1$ & & $255$ & $175$\\
	\end{tabular}
	\caption{\label{LOGTable} $GF(2^8)$ Log and Anti Log Tables}
\end{figure}


\end{document}